\documentclass[twocolumn,amsmath,amssymb]{revtex4}

\usepackage{graphicx}
\usepackage{dcolumn}
\usepackage{bm}
\graphicspath{{./Figures/}}

\def\sqcm{\mbox{ cm$^2$}}

\def\ea0{\mbox{ $ea_0$}}

\def\mF0{\mbox{ $\mu\Phi_0$}}
\def\mF0rtHz{\mbox{ $\mu\Phi_0/\sqrt{\rm Hz}$}}
\def\rtHz{\mbox{$\sqrt{\rm Hz}$}}

\begin{document}

\title{Magnetic susceptibility and magnetization fluctuation measurements of mixed Gadolinium-Yttrium Iron Garnets}

\author{S. Eckel}
\email{stephen.eckel@yale.edu}
\affiliation{Yale University, Department of Physics, P.O. Box
208120, New Haven, CT 06520-8120}

\author{A. O. Sushkov}
\affiliation{Yale University, Department of Physics, P.O. Box
208120, New Haven, CT 06520-8120}

\author{S. K. Lamoreaux}
\affiliation{Yale University, Department of Physics, P.O. Box
208120, New Haven, CT 06520-8120}

\date{\today}

\begin{abstract}
We study the magnetic properties of Gadolinium-Yttrium Iron Garnet
($\text{Gd}_{x}\text{Y}_{3-x}\text{Fe}_5\text{0}_{12}$, $x=3,1.8$) ferrite ceramics. The complex initial permeability is measured in the temperature range 2~K to 295~K at frequency of 1~kHz, and in the frequency range 100~Hz to 200~MHz at temperatures 4~K, 77~K, and 295~K. The magnetic viscosity-induced imaginary part of the permeability is observed at low frequencies. Measurements of the magnetization noise are made at 4~K. Using the fluctuation-dissipation theorem, we find that the observed magnetization fluctuations are consistent with our measurements of the low-frequency imaginary part of permeability. We discuss some implications for proposed precision measurements as well as other possible applications.
\end{abstract}

\maketitle

\section{Introduction}
Rare-earth garnets have generated sustained interest in various soft-ferrite applications, and, more recently, in more fundamental
research. These materials combine attractive magnetic properties (large permeability, low loss angle) with high resistivity, which
leads to their extensive use in microwave applications~\cite{Pardavi-Horvath2000,Zaquine1988}. Most of their applications up to now have been at room temperature, although there are some
more recent suggestions for low-temperature applications also~\cite{Dionne1997}. Gadolinium iron garnet (GdIG), for example, has found
use in adiabatic demagnetization refrigeration (ADR) in satellites~\cite{King2002}. From the perspective of more fundamental
measurements, rare-earth garnets have generated great interest as attractive materials for experiments searching for the parity and
time-reversal invariance-violating permanent electric dipole moment of the
electron~\cite{Lamoreaux2002,Kuenzi2002,Dzuba2002,Mukhamedjanov2003,Heidenreich2005}, as well as for measurements of nuclear anapole
moments~\cite{Mukhamedjanov2005b}, and possibly measurements of the Weinberg angle~\cite{Mukhamedjanov2006}.

The first evidence that GdIG retains its large initial permeability when it is cooled to cryogenic temperatures came from the study
by Pascard~\cite{Pascard1986} of the real part of the magnetic susceptibility versus temperature, spanning the range from
4~K to 564~K (the Curie temperature of GdIG). The complex permeability $\mu = \mu'-i\mu''$ of GdIG was  measured eariler at
radio-frequencies at three temperatures: 300~K, 195~K, and 77~K~\cite{McDuffie1960}. Other properties have also been studied at low
temperature, including the spontaneous and saturation magnetizations~\cite{Geller1965, Wolf1961}, crystal anisotropy~\cite{Pearson1962}, and
ferrimagnetic order~\cite{Yamagishi2005}. However, to our knowledge, there have been no studies of the full complex initial permeability of
Gd-containing ferrites in the temperature range between 4~K and 300~K, and at frequencies up to 1~GHz. Such a study is necessary to evaluate the suitability of these materials for low-temperature applications and precision measurements.

An important emerging application of soft ferrites is magnetic shielding. Sensitive magnetic field measurements have to be performed
inside a set of permeable shields, and are often limited by the magnetic field noise generated by Johnson currents in the shields
themselves~\cite{Allred2002}. Constructing the shields out of high-resistivity soft ferrite materials greatly reduces this source of
magnetic noise at frequencies above approximately 50~Hz~\cite{Kornack2007}. At lower frequencies, however, finite imaginary part of the
permeability generates extra magnetic noise with 1/f power spectrum, in agreement with
the fluctuation-dissipation theorem~\cite{Durin1993,Kornack2007}.

This magnetic noise is related to magnetic viscosity, which arises when the magnetization of a sample is delayed following application of a magnetic field~\cite{Street1994}.  Although initially seen in ferrous materials, this effect has recently been found in nanowires and
other nanostuctures~\cite{Gao2007,Sfikorvanek2001}.  Anomalous magnetic viscosity effects have also been seen in various ferrite materials~\cite{Collocott2002}.  Magnetic viscosity is typically measured by looking at the repsonse of the magnetization after a sudden change in the applied field;
however, it is also characterized by a small $\mu''$ that is independent of frequency for a given magnetic field strength and temperature~\cite{Street1952,Lundgren1981,Prodi1989}.
As shown in the above references, the temperature and field dependences of this effect can be quite complex.

In this work we study the magnetic properties of mixed Gadolinium and Yttrium iron garnets with the chemical formula
$\text{Gd}_{x}\text{Y}_{3-x}\text{Fe}_5\text{0}_{12}$.  Here, $x$ quantifies the ratio of Gadolinium to Ytrrium present
within the garnet.  We present measurements of both real and imaginary parts of the complex initial susceptibility
in the temperature range of 4~K to 300~K, and in the frequency range of 100~Hz to 200~MHz.  We also measure the
magnetization noise of these materials, and confirm the consistency of our results with the fluctuation-dissipation theorem.

\section{Experimental setup}
\label{sec:exp}
\begin{figure}
 \includegraphics[width=\linewidth]{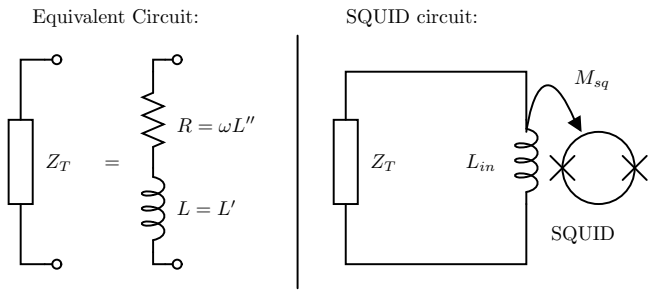}
 \caption{\label{fig:exp} Equivalent circuit for an inductor with a dissipative core (see text) and circuit schematic for the SQUID flux pickup loop, described in Section~\ref{sec:noise}.}
\end{figure}

The inductance of a tightly wound coil on a toroidal core is proportional to the permeability $\mu$ of the core material.  The presence of a complex permeabililty $\mu=\mu'-i\mu''$ creates a complex inductance $L=L'-iL''$.  The impedance for such an toroidal
inductor is $Z_T=i \omega L'+ \omega L''$, which is equivalent to a resistance of $R=\omega L''$ in series with an inductor
$L'$, as shown in Fig.~\ref{fig:exp}. We deduce the real and imaginary parts of the permeability from measurements of the complex impedance of such a toroid.

Two toroidal samples: one pure GdIG ($\text{Gd}_{3}\text{Fe}_5\text{0}_{12}$), and one mixed GdYIG ($\text{Gd}_{1.8}\text{Y}_{1.2}\text{Fe}_5\text{0}_{12}$), were purchased from Pacific Ceramics. These were manufactured using their standard process.  Powder
was granulated and screened, using 99.99\% pure Gd, 99.999\% pure Y, and 99.6\% pure Fe oxides.  The typical grain size was 5-10
microns.  The samples were pressed and fired at approximately $1400^{\circ}$C for 15 hours.  The toroids were made with square
cross sections, with inner radius of 0.5~cm, outer radius of 1.5~cm, and a height of 1~cm.

To measure the real part of the permeability, $\mu'$, at frequencies less than 1~MHz, the toroidal samples were wrapped with approximately 30 turns of wire, and the resulting self-inductance was measured with a QuadTech Series 1200 LCR meter. This method does not give reliable measurements for the imaginary part of the permeability, however, since the ohmic resistance of the winding dominates the loss due to $\mu''$. For measurements of $\mu''$ at frequencies less than 1~MHz, the toroidal samples were wrapped with a primary winding of approximately 35 turns, and a secondary winding of approximately 80 turns. Litz wire was used to reduce skin effects. The complex mutual inductance of the two windings was then measured by a four-point measurement with the LCR meter. Since the input impedance of the voltage inputs on the meter is large ($> 10\ \text{M}\Omega$), the current flowing in the secondary winding is very small, thus the ohmic resistance of the secondary does not affect the measurement of the imaginary part of the susceptibility (on the level of a fraction of one percent). The inter-winding capacitance, however, limits this technique to frequencies below 100~kHz, at higher frequencies the small impedance of the shunt capacitor results in a substantial current in the secondary winding.

A Hewlett-Packard series 4100A impedance analyzer was used for measurements in the range of 1~MHz to 200~MHz.  The sample was placed in a toroidal-shaped copper housing to reduce the number of turns around the toroid to one.  This
minimized the electrical length of the wrapping such that it was always less than the wavelength. However, it introduced
capacitive effects which prevented reliable measurements at frequencies above 200~MHz.  The self inductance of the single turn inductor was measured with the impedance analyzer.
Ohmic resistance effects lead to errors on the order of 1\%, since the series impedance due to $\mu''$ dominates at such high frequencies.

For all the inductance measurements, the strength of the applied magnetic field was no more than $H_0\sim3\ \text{A/m}$, estimated from the current flowing in the toroid windings. This is much smaller than the ferrite's saturation field ($H_0\sim10^5\ \text{A/m}$)~\cite{Yamagishi2005}, which ensures that we always measure the initial susceptibility of the material.

Low-temperature measurements were performed with the samples mounted in a G-10 holder inside a Janis model 10CNDT cryostat. Temperature was measured with a LakeShore model DT-670C-SD silicon diode temperature sensor, read out by a LakeShore Model 325 temperature controller, with accuracy of 0.1~K.   At 4~K, temperature-dependent effects on the calibration of the QuadTech LCR meter were corrected by making a measurement of $\tan\delta=\mu''/\mu'$ using a SQUID magnetometer.  Multiple-turn coils were wrapped around toroidal non-magnetic (G10) and GdIG samples and the coils were connected in series.  A single-turn superconducting pickup loop was wrapped about each sample and each pickup loop was connected to a SQUID.  An sinusoidal current was passed through the coils, and the phase shift between the non-mangetic and GdIG samples was recorded.  The $\tan\delta$ recorded by the SQUIDs was then used to correct the phase-shift error in the calibration of the QuadTech LCR meter at 4~K.  More details on the SQUID setup can be found in Sec.~\ref{sec:noise} and Ref.~\cite{EDMPaper}.  For temperature dependent measurments, temperature variation was accomplished by removing the cryogens and letting the cryostat warm up, while a control computer read out the temperature controller and the LCR meter. Temperature dependent measurements were typically separated by 0.1~K, and the total warming time was approximately 12 hours.

%
%
%

\section{Initial Susceptibility}
\label{sec:susc}
\begin{figure}
 \includegraphics[width=\linewidth]{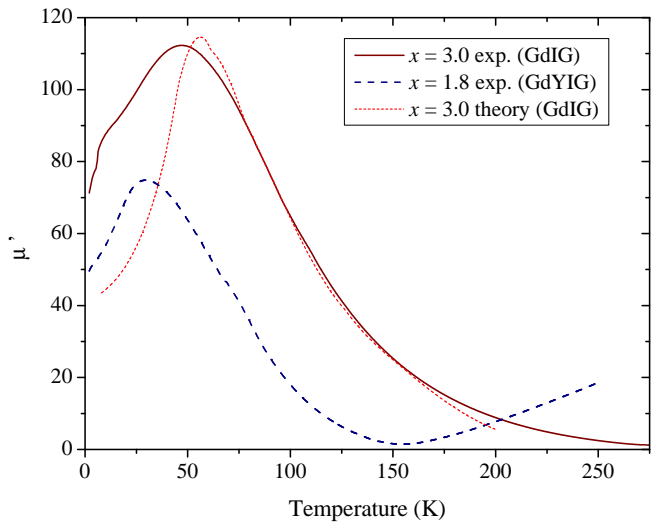}
 \caption{\label{fig:susvst} The real part of the initial permeability of GdIG and GdYIG versus temperature.
The Globus model (Eq.~\ref{eq:globusmodel}) described in the text is shown in fine dashed red as the best fit to the data between
65 K and 150 K.  All data were obtained at a frequency of 1 kHz and $H\approx3$ A/m applied.}
\end{figure}

The measured real part of permeability, $\mu'$, as a function of temperature is shown in Fig.~\ref{fig:susvst} for GdIG and GdYIG ferrites. It is evident that both these ferrites retain their large permeability down to 4~K.
The magnetic properties of mixed Gadolinium-Yttrium iron garnets are well described by a model with three magnetic sublattices: the c-sites containing Gd$^{3+}$ ions, the a-sites containing Fe$^{3+}$ ions, and the d-sites containing Fe$^{3+}$ ions~\cite{Geller1961}. The Y$^{3+}$ ions are not magnetic, they substitute for Gd$^{3+}$. At the compensation temperature, the magnetizations of the three sublattices compensate each other, so that the net magnetization vanishes, and the permeability approaches unity~\cite{kittel}. As evident from Fig.~\ref{fig:susvst}, the compensation temperature for $\text{Gd}_{1.8}\text{Y}_{1.2}\text{Fe}_5\text{0}_{12}$ is 155~K, in good agreement with measurements in Ref.~\cite{Yamagishi2005}. The compensation temperature for pure GdIG is 290 K~\cite{Rodic1990,Yamagishi2005,Geller1965}, which is not visible in the plot.

As claimed in Ref.~\cite{Pascard1986}, the model of domain wall bulging due to Globus~\cite{Globus1971,Liorzou2000} provides a
satisfactory explanation of the temperature dependence of $\mu'$ for GdIG ($x=3.0$).  In this model
\begin{equation}
\label{eq:globusmodel} \mu'-1=\frac{3\pi M_s^2}{4 l |K_1|}D_m\ ,
\end{equation}
where $M_s$ is the saturation magnetization, $K_1$ is the magnetocrystalline anisotropy, $D_m$ is the mean grain size, and $l$ is
a constant, with dimensions of length, that depends on the properties of the material.  Taking the data for $M_s$ and $K_1$ from Refs.~\cite{Geller1965,Pearson1962} and fitting in the least squares
sense the constant $l$ yields the theoretical fit in Fig.~\ref{fig:susvst}.  There is good agreement with our data between 65 K and 200 K.
Lack of agreement below 50 K, however, suggests that the single-crystal
samples used in Refs.~\cite{Geller1965,Pearson1962} do not have the same low-temperature properties as the pressed ceramic used for the present work.

Our measurements indicate the maximum permeability of 112 at 47~K for GdIG. Susceptibility data from Ref.~\cite{Pascard1986}, however, shows a peak in the GdIG susceptibility of $\mu'\approx70$ at 56~K.  This discrepancy can be accounted for within the model (\ref{eq:globusmodel}) by the difference in grain size, $D_m$, of the ceramic samples: the samples used in Ref.~\cite{Pascard1986} have $D_m\sim4\ \mu$m, whereas our samples have $D_m\sim7\ \mu$m.

\begin{figure}
 \includegraphics[width=\linewidth]{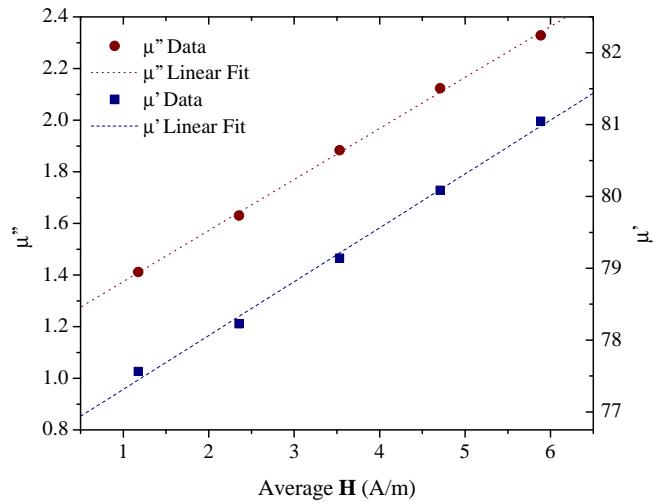}
 \caption{\label{fig:susvsh} The real and imaginary parts of permeability of GdIG versus applied field $\mathbf{H}$, averaged over the volume of the sample.  This data was obtained at a frequency of 1 kHz and temperature of 4 K.}
\end{figure}
At fixed temperature and frequency, the complex susceptibility depends on the applied field $H$. The measurements of $\mu'$ and $\mu''$ for various applied fields $H$ are shown in Fig.~\ref{fig:susvsh}.  We always observed linear dependence of the magnetic susceptibility on $H$, implying that the magnetization is given by $M=\chi H + \frac{1}{2}\alpha H^2$.  This nonlinear dependence of magnetization on applied field leads to the low-field Rayleigh hysteresis loop~\cite{Chikazumi1997}. Our four-point measurements of $\mu''$ at low frequency ($f<100$ kHz) were always made at several applied fields $H$ and linearly extrapolated to zero, as shown in Fig.~\ref{fig:susvsh}.

\begin{figure}
 \includegraphics[width=\linewidth]{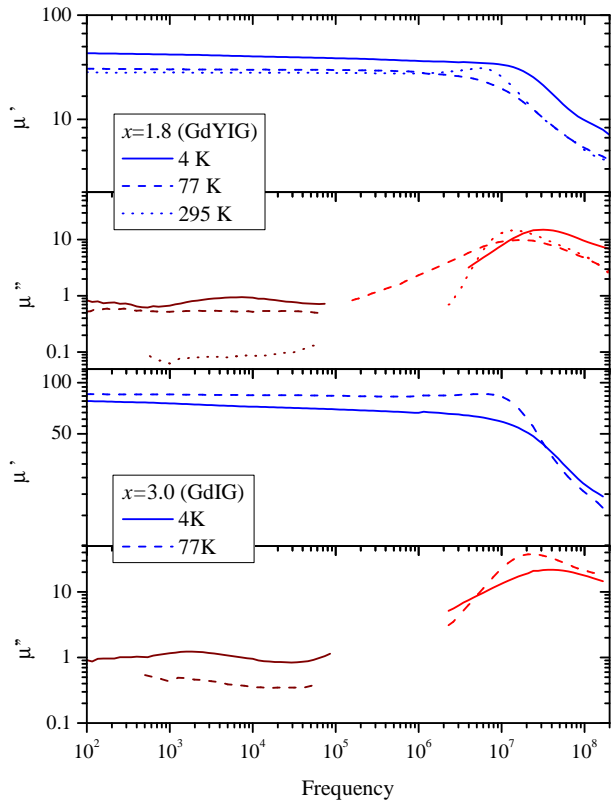}
 \caption{\label{fig:susvsf} The real and imaginary parts of permeability of GdIG and GdYIG versus frequency.
 Since $\mu'\sim1$ for GdIG at room temperature, no data is shown.  All
data were obtained with an applied field of $H\approx3$ A/m.  Low frequency ($f<100$ kHz) $\mu''$ data were obtained by a
four point measurement and extrapolated to zero applied field, as described in the text.}
\end{figure}
The measurements of the complex susceptibility as a function of frequency at 4~K, 77~K, and 295~K are shown in Fig.~\ref{fig:susvsf}. No data are shown for pure GdIG at 295~K, because this is close to its compensation temperature, so the sample was very nearly non-magnetic at this temperature. It is evident from the data that near 20~MHz the permeability of both samples exhibits a resonance. This ``natural resonance'' occurs at the frequency $\omega_r=\gamma H^A$, where $\gamma$ is the gyromagnetic ratio, and $H^A\propto K_1/M_s$ is the crystalline anisotropy field~\cite{Chikazumi1997,Smit1959}. Using Eq.~(\ref{eq:globusmodel}) we can re-write the resonance frequency as:
\begin{equation}
\omega_r\propto\frac{M_s}{\mu'-1}\ .
\end{equation}
Using the data for saturation magnetization in Ref.~\cite{Geller1965}, and our data for $\mu'-1$, it is apparent that as the temperature decreases, $M_s$ grows faster than $\mu'-1$, thus the natural resonance should shift to higher frequency with lower
temperature, which is indeed the case as can be seen from the data in Fig.~\ref{fig:susvsf}. A similar trend is observed in Ref.~\cite{McDuffie1960}.

In spite of the shift of the natural resonance to higher frequency, the low-frequency imaginary part of permeability increases as the temperature is lowered. This, as well as inspection of the data in Fig.~\ref{fig:susvsf}, shows that the frequency dependence of the permeability is not as simple as the models suggested in Ref.~\cite{Budker2006}, for example. It appears that at low frequencies ($f<100$~kHz) the magnitude of $\mu''$ is dominated not by the wing of the natural resonance, but by some frequency-independent dissipation mechanism. This is known as magnetic viscosity.

The presence of a non-zero $\mu''$ is indicative of a phase shift of magnetization with respect to the applied field $H$.  The
phase lag, or loss angle, $\delta$ is given by $\tan\delta=\mu''/\mu'$.  For a given
temperature, and at frequencies much lower than the natural resonance, the loss angle is empirically parameterized by the lag equation~\cite{Bozorth1951}:
\begin{equation}
\label{eq:legg} 2\pi\frac{\tan\delta}{\mu'}=c + a H + e f,
\end{equation}
where $c$, $a$, and $e$ are parameters, and $f$ is the frequency of the applied field $H$, which is much less than the saturation field.
The last term accounts for eddy current effects. However, in the present case, eddy current losses are
negligible since our ferrite samples are very good insulators. The second term describes hysteresis effects described above, thus the parameter $a$ is a measure of non-linearity in the system.  The parameter $c$ is a constant allowing for frequency-independent loss at zero applied field, this is the magnetic viscosity.  Using the low-frequency measurements of $\mu''$ versus frequency and applied field $H$, values of $a$ and $c$ were extracted at 4 K, 77 K, and 295 K.  These values are shown in Table~\ref{tab:losses}. Once again, we do not give any results for pure GdIG ($x=3$) at 295~K, since this is very close to its compensation temperature.

\begin{table}
\caption{\label{tab:losses}Parameters of the loss equation (\ref{eq:legg}) for GdIG and mixed GdYIG at various temperatures.  Values were compiled by least squares fit to all available data less than 10 kHz.  Error bars are extracted from the fit ($1\sigma$).} \begin{ruledtabular} \begin{tabular}{lccr} $x$ & $T$ & $c$ & $a$ (A/m)$^{-1}$ \\ \hline 3.0 & 77 K & $(3.51\pm0.40)\times10^{-4}$ & $(2.73\pm0.13)\times10{-4}$ \\  & 4 K & $(1.14\pm0.14)\times10^{-3}$ & $(2.30\pm0.48)\times10^{-4}$ \\ \hline
1.8 & 295 K & $(5.75\pm0.44)\times10^{-4}$ & $(3.68 \pm0.18)\times10^{-4}$ \\  & 77 K & $(2.93\pm0.12)\times10^{-3}$ & $(6.75\pm0.31)\times10^{-4}$ \\  & 4 K & $(2.05\pm0.30)\times10^{-3}$ & $(6.73\pm1.13)\times10^{-4}$ \\ \end{tabular} \end{ruledtabular} \end{table}


Magnetic viscosity is characterized by time dependent change in the magnetization after a change in the applied magnetic field. When a small magnetic field is applied to a ferromagnet, magnetization changes primarily due to domain wall movement. As it moves, the domain wall traverses a complex potential energy landscape~\cite{Chikazumi1997}. The time scale of the thermally-driven movement across a potential energy barrier of height $E_d$ is given by an Arrhenius-type equation: $\tau=\tau_0\exp(E_d/k_BT)$~\cite{Smit1959}. In a macroscopic sample, there are multiple domain walls, each seeing some distribution of such activation energies. If this distribution is flat over a wide range of potential barrier heights, then $\mu''$ is independent of frequency, and the magnetic viscosity term $c$ appears in the expression (\ref{eq:legg}) for the lag angle~\cite{Street1949,Street1952,Lundgren1981,Prodi1989}.

\begin{figure}
 \includegraphics[width=\linewidth]{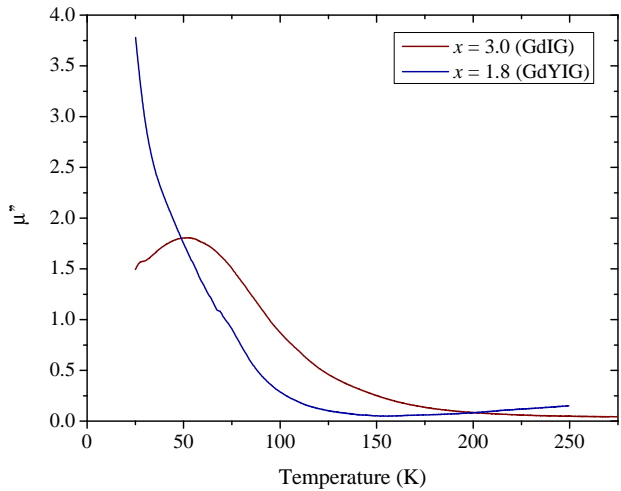}
 \caption{\label{fig:immuvsT} Measurement of $\mu''$ versus temperature for GdIG and GdYIG.
All data were obtained at a frequency of 1 kHz and $H\sim3$ A/m applied.  Data below 25~K was unreliable due to significant temperature dependent effects on the QuadTech LCR meter's calibration.}
\end{figure}
The temperature dependence of magnetic viscosity of our GdYIG samples can be inferred from Fig.~\ref{fig:immuvsT}, where $\mu''$ is plotted as a function of temperature. There are several competing effects that lead to a complicated temperature dependence~\cite{Street1952}. On the one hand, at low temperature, the domain wall relaxation times grow longer, increasing the magnitude of $\mu''$. On the other hand, as the temperature is lowered even further, some of the domain walls become pinned, no longer contributing to the initial susceptibility, decreasing the magnitudes of both $\mu'$ and $\mu''$, as seen in the $x=3.0$ curve in Figs.~\ref{fig:immuvsT} and~\ref{fig:susvst}.

\section{Magnetization Noise}
\label{sec:noise}

The magnitude of the magnetization noise is the key property of the GdYIG garnet ferrites that determines the feasibility of their application in magnetic shielding as well as in experiments searching for parity and/or time-reversal symmetry violations. We measured the magnetization noise of our samples and compared the results with our data for the complex initial permeability using the fluctuation-dissipation theorem.

The noise measurements were made at 4~K inside the cryogenic dewar described in Section~\ref{sec:exp}. A Quantum Design model 50 superconducting quantum interference device (SQUID) magnetometer was connected to the sample under study by a niobium wire pickup loop. The one-turn loop was wound around the square cross-section of the toroidal sample, so that the total flux pickup area was 1~cm$^2$. The SQUID control and readout was performed by the control unit (model 5000), connected to the SQUID via  a 4-meter MicroPREAMP cable. The SQUID and the sample were mounted on G-10 holders, and enclosed in superconducting shielding, made of 0.1-mm thick Pb foil, glued to the inner surfaces of two G-10 cylinders. The magnetic field shielding factor of our setup is greater than $10^9$. Some more details on our noise measurement setup are given in Ref.~\cite{EDMPaper}.

\begin{figure}
 \includegraphics[width=\columnwidth]{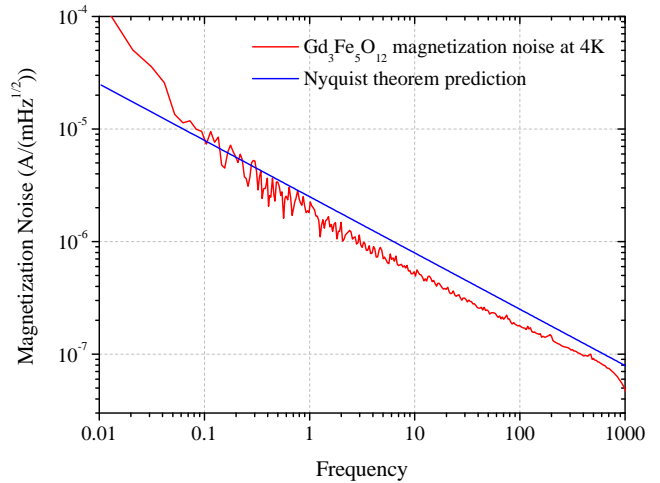}
 \caption{\label{fig:noise} Magnetization noise of GdIG at 4.2 K.  The blue line is the result of using the fluctuation-dissipation theorem, Eq.~(\ref{eq:bnoise}), with the value for the imaginary part of permeability $\mu'' = 1.1$, taken from measurements shown in Fig~\ref{fig:susvsf}.  The SQUID white noise level corresponds to $10\mbox{ fT}/\rtHz$ on this plot.}
\end{figure}

The measured magnetization noise amplitude spectrum for a pure GdIG sample is shown in Fig.~\ref{fig:noise}. The noise data for the mixed GdYIG sample is within 40\% of the pure GdIG noise; we do not show it, since the two curves would lie on top of each other, given the plot's logarithmic scale. The measurement bandwidth of 1~kHz is set by the output analog filter to prevent aliasing with our sampling rate of 12~kHz. The SQUID magnetometer intrinsic sensitivity in this frequency range is 7~fT/\rtHz; within our bandwidth, the magnetization noise is always above the magnetometer sensitivity.

According to the fluctuation-dissipation theorem, any dissipative material exhibits thermal noise. We measured the dissipation of our ferrite samples in Section~\ref{sec:susc}, this is given by the imaginary part $\mu''$ of the complex permeability. We can therefore check that our measurements of the magnetization noise and the dissipation are consistent using the fluctuation-dissipation theorem. Consider the circuit shown in Fig.~\ref{fig:exp},
where the complex impedance $Z_T$ of the pickup loop around the permeable sample is in series with the input inductance $L_{in}$ of the SQUID.  The real part of $Z_T$, $\omega L''$, is a source of voltage noise, whose spectral density is given by the Nyquist theorem:
\begin{equation}
\label{eq:nyquist} (\delta V)_\omega=\sqrt{4k_B T \omega L''},
\end{equation}
where $k_B$ is Boltzmann's constant, $T$ is temperature, $\omega = 2\pi f$ is angular frequency, and $L''$ is the imaginary part of the pickup loop inductance. The current noise in the circuit can be obtained by dividing Eq.~(\ref{eq:nyquist}) by the total impedance, $Z_T+\omega L_{in}\approx\omega (L'+L_{in})$. This current couples to the SQUID through the mutual inductance $M_{sq}$. The resulting flux through the SQUID is the same as would be created by a lossless pickup loop around the sample with magnetization noise given by:
\begin{equation}
\label{eq:bnoise} (\delta M)_\omega=\frac{1}{\mu_0 A}\sqrt{\frac{k_B T L''}{2 \pi f}}\ ,
\end{equation}
where $A=1.0\sqcm$ is the area of the pickup loop and $\mu_0$ is the permeability of free space.  The dependence of the single turn pickup loop's inductance on $\chi=\mu-1$ was found empirically and is described well by a linear expansion, e.g. $L=L_0+\chi L_1$, where $L_0$ and $L_1$ are the expansion coeffecients. Since these coffecients are real, the imaginary part of the inductance is  given by $L''=\chi'' L_1=\mu'' L$.

Using Eq.~(\ref{eq:bnoise}) and our empirical determination of $L''$, we can compare our data for $\mu''$ of GdIG at 4~K with our measurement of its magnetization noise. From the GdIG data shown in Fig.~\ref{fig:susvsf} we can deduce, with good accuracy, a constant $\mu'' = 1.1$ for frequencies below 1~kHz. With this value, Eq.~(\ref{eq:bnoise}) gives a prediction for the magnetization noise, shown as the blue line in Fig.~\ref{fig:noise}. Note the $1/\sqrt{f}$ dependence characteristic of magnetic viscosity. The agreement with experimental noise measurements is remarkable, given that $\mu''$ was measured with an entirely different method.

\section{Conclusions}
We have measured the full complex permeability of two GdYIG ceramics in the temperature range 2~K to 295~K and confirmed that the real part of the permeability remains large (78 and 52 respectively) at 4~K. The Globus magnetization model fits the results well between 65~K and 200~K. We also measured the full complex permeability as a function of frequency in the range 100~Hz to 200~MHz at temperatures of 4~K, 77~K, and 295~K. We observed the natural resonances near 20~MHz, and detected the frequency-independent imaginary permeability due to magnetic viscosity effects for frequencies less than 100~kHz: $\mu''=1.1$ for GdIG and $\mu''=0.8$ for GdYIG at 4~K. To our knowledge, such effects have not been studied for rare-earth ferrites in such a wide temperature range.
Using the fluctuation-dissipation theorem, these results were compared with direct measurements of the magnetization noise, with very good agreement.

Our noise measurements at 4~K show that the intrinsic $1/\sqrt{f}$ magnetization noise in GdIG and GdYIG is larger than the SQUID magnetometer noise for frequencies below approximately 100~kHz. This implies that precision measurements involving these materials either have to be carried out at higher frequencies, to avoid this noise, or have to make use of the geometry-dependent demagnetizing fields to reduce the coupling of the noise to the measurement device~\cite{EDMPaper}.

A possible application of the materials studied in the present work is to magnetic shielding of high-T$_c$ SQUIDs at 77~K. For a cylindrical shield of radius 5~cm and thickness 1~cm (the dimensions in Ref.~\cite{Kornack2007}), made of GdIG, the shielding factor is 6. The magnetic field noise on the axis can be predicted from our measurement of $\mu''=0.5$ at 77~K, resulting in $\delta B = 10$~fT/$\rtHz$ at 1~Hz. This is at the level of a high-T$_c$ SQUID's sensitivity.  Such a shield can be used inside a set of $\mu$-metal shields to screen external magnetic fields and the Johnson noise due to the innermost layer of the $\mu$-metal shields, without generating excessive Johnson noise itself.

\begin{acknowledgments}
The authors would like to acknowledge useful discussions with Woo-Joong Kim and Dmitry Budker. This work was supported by Yale University.
\end{acknowledgments}

\bibliography{GdIGLibrary,EDMLibrary}

\end{document}